\documentclass[floatfix,showkeys,twocolumn,citeautoscript,showpacs,amsmath,amssymb,epsf,aps]{revtex4}
\usepackage{dcolumn}
\usepackage{epsfig}

\usepackage{graphicx}% Include figure files
\usepackage{longtable}% Include figure files
\usepackage{dcolumn}% Align table columns on decimal point
\usepackage{bm}% bold math

\def\etal{{\em et al. }}
\def\BNF{B$_{24}$N$_{24}$ }
\def\ocm{cm$^{-1}$ }

\def\BNE{B$_{28}$N$_{28}$ }

\def\BNS{B$_{36}$N$_{36}$ }

\begin{document}
\preprint{USC/002}

\title{ Theoretical   infra-red,  Raman, and Optical spectra of the \BNS cage}

\author{Rajendra R. Zope$^{1}$, Tunna Baruah$^{2,3}$, Mark R. Pederson${^2}$, and Brett I. Dunlap$^4$}

\affiliation{$^1$Department of Chemistry, George Washington University, Washington DC, 20052, USA}

\affiliation{$^2$Code 6392, Center for Computational Materials Science, US Naval Research Laboratory,
 Washington DC 20375, USA}

\affiliation{$^3$Department of Electrical Engineering and Materials Science Research Center, 
Howard University, Washington DC 200059, USA}

\affiliation{$^4$Code 6189, Theoretical Chemistry Section, US Naval Research Laboratory,
Washington, DC 20375}

\date{\today}

\begin{abstract}
  The \BNS fullerene-like cage structure was proposed as candidate structure for the single-shell
boron-nitride cages observed in electron-beam irradiation experiment.  We have performed  all 
electron density functional calculations, with large polarized Gaussian basis sets, on the \BNS
cage. We show that the cage is energetically and vibrationally stable. The infra-red,
Raman and optical spectra are calculated. The predicted spectra, in combination with experimentally 
measured spectra, will be useful in conclusive assignment of the proposed \BNS cage. 
The vertical and adiabatic ionization potentials as well as static dipole polarizability 
are also reported.
\end{abstract}

\pacs{ }

%Use showkeys class option if keyword display desired

\maketitle
     The discovery of carbon fullerenes\cite{KHOCS85}, the synthesis of 
boron nitride (BN) nanotubes\cite{Zettl}, and the fact that the BN pair 
is isoelectronic with a 
pair carbon atoms, led to  the search for fullerene analogues of BN.  The carbon 
fullerenes are made close by introducing 12 pentagons in the hexagonal network 
of carbon atoms. The exact BN analogues of fullerenes are not preferred as the presence of 
pentagonal ring does not permit alternate sequence of B and N atoms. However, by 
Euler's theorem, it can be shown that fullerene-like alternate BN cages can 
be formed using six isolated squares\cite{SSL95,ZSK97}. The square contains 
four BN bonds with alternate boron and nitrogen atoms.
Several theoretical works have reported the possible candidate 
structures for the BN cages\cite{SSL95, ZSK97, Strout,P2000,WJ04,ZD04,ZBPD04,Oku00}.  

On the experimental side, there are also a few reports of synthesis of hollow BN 
structures\cite{Stephan98,Goldberg98,Oku00,Oku03,Oku04}. In a series of experiments 
by Oku and coorkwers, boron nitride clusters of different sizes were 
reported\cite{Oku00,Oku03,Oku04}. Perhaps the most interesting result of these works was 
the production of \BNF in abundance. We performed theoretical calculations\cite{ZD04,ZBPD04} 
on \BNF, and think that the roundness of the octahedral \BNF cluster is the most likely
explanation of its unexpected abundance.   

In another set of experiments,
 St\'ephan \etal irradiated  BN-powder and BN material\cite{Stephan98} and
observed small BN cage-like molecules.  The irradiated derivatives were either 
close-packed cages or nested cages. The most observed cages in the experiment 
had diameters in the size range 4 to 7 \AA; the octahedral B$_{12}$N$_{12}$, B$_{16}$N$_{16}$,
and \BNE cages were proposed as possible structures.  
Subsequent tilting experiment in an electron microscope 
by  Goldberg and coworkers permitted viewing of the irradiation-induced BN cages in 
different directions and corroborated the proposal of octahedral structures for the observed
BN cages\cite{Goldberg98}. From the  high resolution transmission electron 
microscopy images (HRTEM), they observed  single-shelled cages of size  9-10 \AA,
with rectangle-like outlines. The \BNS cage has roughly similar dimension and was proposed
as candidate structure by Alexandre \etal \cite{AMC99} and Oku \etal \cite{Oku00}. 
Alexander \etal also performed a finite-range pseudopotential density functional calculation 
on the \BNS and showed that the \BNS cage is energetically stable
and has large energy gap between its  highest molecular orbital and lowest molecular orbital
(HOMO-LUMO). These conclusions were subsequently confirmed by all electron calculations using
the Slater-Roothan (SR) method\cite{ZD04}. These calculations show that the proposed \BNS
cage has overall size of about 8 \AA, smaller than single-shelled BN cages observed 
in the experiment.
So far the proposal of \BNS as the observed structure has not yet been
verified in alternative experiments. It is the purpose of the present work to ascertain 
the stability of proposed cage structure of \BNS with respect to small distortions and to 
provide its spectroscopic data : the infra-red (IR) spectra, Raman and optical spectra, for 
comparison  with experimental spectra.  The comparison of the calculated and experimentally
measured  spectra has played an important role  in structure 
determination in past.  We hope that the present work will stimulate further experimental works,
which combined with the predicted spectra, would unambiguously confirm the 
proposal of \BNS fullerene-like cage.

\begin{table}
\caption{The optimized coordinates of ineqivalent atoms in \BNS.}
\begin{tabular}{lccc}
\hline
Atom  &   X         &     Y          &    Z       \\
\hline
B  &    -1.340   &    4.713    &   4.713  \\
B  &    -7.771   &   -1.224    &  -1.224  \\
B  &     6.099   &    2.739    &   2.739  \\
N  &     1.457   &    4.876    &   4.876  \\
N  &     8.189   &    1.504    &   1.504  \\
N  &    -6.057   &   -2.702    &  -2.702  \\
\hline
\end{tabular}
\label{tab:xyz}
\end{table}

 Our calculations were carried out at the all electron level using the { NRLMOL} suite of 
codes \cite{NRLMOL}. The exchange-correlation effects were treated at the level of  the
generalized gradient approximation using the Perdew-Burke-Ernzerhof scheme(PBE-GGA)\cite{PBE}.
Large polarized Gaussian basis sets\cite{Porezag99} optimized for density functional 
calculations (PBE-GGA) are employed. The B-atom basis set consists of  5 $s-$type,
4  $p-$type, and 3 $d-$type  orbitals each of which were constructed from a linear combination of  
12 primitive Gaussians. For N atoms, 5 $s-$type, 4 $p-$type, and 3 $d-$type functions comprised 
of 13 Gaussian are used. The geometry optimization was performed using the limited-memory
Broyden-Fletcher-Goldfarb-Shanno scheme until the forces on each atom were less 
than 0.001 a.u.  The self-consistent calculation was iterated until the energy 
difference of successive iteration was smaller 0.0000001 a.u. The vibrational 
frequencies were calculated by diagonalizing  a dynamical matrix constructed by displacing 
the atoms by small amount\cite{Porezag96}. The IR and Raman frequencies were determined 
from the derivative of the dipole moment and the polarizability tensor.

\begin{figure}
\epsfig{file=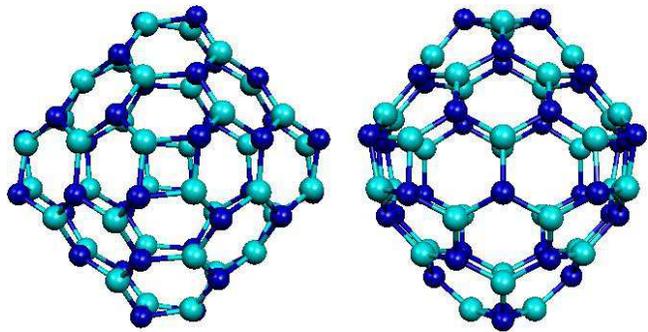,width=8.5cm,clip=true}
\caption{\label{fig1} (Color online) Two different views of  optimized 
\BNS cages. In the second view the \BNS cage is rotated by 45$^o$ around a  vertical
axis passing through center of square.}
\end{figure}

\begin{figure}
\epsfig{file=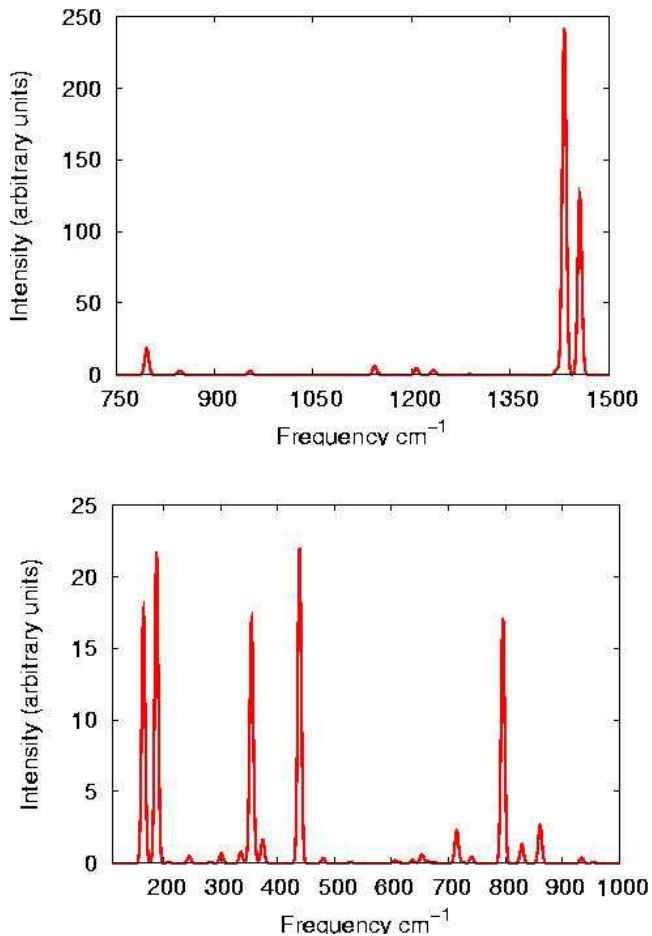,width=8.5cm,clip=true}
\caption{\label{fig:ir} (Color online) The infra-red (top) and Raman (lower panel) spectrum of \BNS cage.}
\end{figure}

%\begin{figure}
%\epsfig{file=raman.eps,width=8.5cm,clip=true}
%\caption{\label{fig:raman} (Color online) Raman spectrum of \BNS cage.}
%\end{figure}

   The optimization of \BNS cage (See Fig. \ref{fig1}) was started using its optimized geometry obtained by 
the SR  method\cite{ZD04}.
%     The optimized \BNS cage is shown in Figure \ref{fig1}. 
The cage has 
$T_d$ symmetry and can be generated from six inequivalent atoms, three for boron, and 
three for nitrogen, using symmetry operations. The optimized coordinates of the 
inequivalent atoms are given in Table \ref{tab:xyz}. These coordinates are with respect 
to the origin at the center of mass.  The average spherical radius of 
the cage, obtained as the mean of distance of each atom from the center of mass, is
3.94 \AA.  The SR method predicts a cage of 3.76 \AA \, radius\cite{ZD04}.
Thus, the cage has roughly dimension of 8 \AA, somewhat smaller than the 
observed dimensions of cage structure in the experiment.   Part of this discrepancy could 
be due to the thermal effects as our calculations are at $T=0K$. 
  The cage is energetically 
stable with respect to isolated atoms and has a binding energy (BE) of 8.43 eV per
atom.  
The SR method gives a binding energy of 7.53 eV/atom using the 6-311G* basis.
Th present BE is higher than that of \BNF cages, which have energies of about
8.3 eV\cite{ZBPD04}.  This result agrees with our predictions by 
SR method\cite{ZD04}. 
The trend of larger BN cages being energetically more stable
than smaller ones is similar to that observed in case of carbon fullerenes\cite{D91}.
The present calculations, consistent with earlier predictions by 
lower theories, show that the \BNS cage is characterized by a large HOMO-LUMO gap of 5.0 eV.
The HOMO level has t$_2$ symmetry.   
The vertical ionization potential (VIP) is  calculated as the self-consistent energy 
differences of the cage and its positive ion with the same ionic configurations.
The calculated VIP is  8.2  eV. This agrees well with 8.4 eV, obtained by the SR method\cite{ZD04}.
We have also determined the adiabatic ionization potential
from the energy differences of the neutral \BNS cage and its optimized singly charged cation.
The adiabatic ionization potential is   8.1 eV, which is an overestimate  as 
the cation was optimized under symmetry constraint. 
The mean polarizability  of \BNS calculated from the finite field method is 78 \AA$^3$. 
The nuclear frame of \BNS was assumed to be frozen during the polarizability calculation.

The vibrational analysis within the harmonic approximation indicate that all 
frequencies are real.  Thus, the \BNS cage is vibrationally stable and corresponds 
to a minimum on the potential surface.   The  calculated IR spectrum is presented 
in Fig. ~\ref{fig:ir}.  The IR spectrum is broadened by 6 \ocm \!\!.  The IR spectra 
shows two conspicuous peaks at  1432 \ocm and 1456 \ocm ; the intensity of the 
latter peak (42  Debye$^2$/amu/\AA$^2$) is half of the former peak. A weak 
but noticeable absorption (intensity 6  Debye$^2$/amu/\AA$^2$)  is found at 797 \ocm.
All the peaks have  $T_2$ symmetry.  The Raman spectrum 
is also shown in Fig. ~\ref{fig:ir}.  The Raman spectrum is characterized by multiple
peaks in the frequency range $150-900$ \ocm.   Of the two prominent peaks in the low frequency 
range, the first one is at 165 \ocm and is of $E$ type. The second one at 188 \ocm 
is due to the triply degenerate mode of $T_2$ symmetry. All other Raman active frequencies: 
at 354, 439, 796, 847, and  869 \ocm are of $A_1$ symmetry.   The absorption at 354 \ocm 
corresponds to symmetric stretching or breathing mode. All vibrational frequencies 
along with their symmetry type are listed in Table \ref{vibs}.

\begin{table*}[h]
\caption{The frequencies, symmetry, IR and Raman activeness of the vibrational modes.}
\label{vibs}
\begin{ruledtabular}
\begin{tabular}{lccccccc}
 Frequency        & Symmetry & IR    &  Raman    & \hskip 0.7in   Frequency                  &  Symmetry & IR     & Raman  \\ 
 (\ocm)           &          &active &  active   & \hskip 0.7in   (\ocm)                     &           & active &  active       \\
\hline   
     164   &  E      &        & $\surd$   & \hskip 0.7in       796     &  T$_2$  &$\surd$ & $\surd$    \\
     187   &  T$_2$  &$\surd$ & $\surd$   & \hskip 0.7in       806     &  A$_2$  &        &            \\
     208   &  A$_1$  &        & $\surd$   & \hskip 0.7in       829     &  A$_1$  &        & $\surd$    \\
     222   &  T$_1$  &        &           & \hskip 0.7in       847     &  T$_2$  &$\surd$ & $\surd$    \\
     245   &  T$_2$  &$\surd$ & $\surd$   & \hskip 0.7in       860     &  A$_1$  &        & $\surd$    \\
     269   &  T$_1$  &        &           & \hskip 0.7in       905     &  T$_1$  &        &            \\
     282   &  E      &        & $\surd$   & \hskip 0.7in       933     &  E      &        & $\surd$    \\
     293   &  T$_1$  &        &           & \hskip 0.7in       953     &  T$_2$  &$\surd$ & $\surd$    \\
     301   &  T$_2$  &$\surd$ & $\surd$   & \hskip 0.7in       966     &  E      &        & $\surd$    \\
     335   &  E      &        & $\surd$   & \hskip 0.7in       967     &  T$_1$  &        &            \\
     346   &  T$_2$  &$\surd$ & $\surd$   & \hskip 0.7in       976     &  A$_2$  &        &            \\
     354   &  A$_1$  &        & $\surd$   & \hskip 0.7in       979     &  T$_1$  &        &            \\
     360   &  E      &        & $\surd$   & \hskip 0.7in      1006     &  T$_2$  &$\surd$ & $\surd$    \\
     369   &  T$_1$  &        &            & \hskip 0.7in      1029     &  A$_1$  &        & $\surd$    \\
     374   &  T$_2$  &$\surd$  & $\surd$   & \hskip 0.7in      1061     &  T$_2$  &$\surd$ & $\surd$    \\
     419   &  T$_1$  &         &           & \hskip 0.7in      1078     &  E      &        & $\surd$    \\
     437   &  T$_2$  &$\surd$  & $\surd$   & \hskip 0.7in      1140     &  T$_1$  &        &            \\
     438   &  E      &         & $\surd$   & \hskip 0.7in      1143     &  T$_2$  &$\surd$ &            \\
     462   &  A$_2$  &         &           & \hskip 0.7in      1154     &  E      &        & $\surd$    \\
     480   &  T$_2$  &$\surd$  & $\surd$   & \hskip 0.7in      1157     &  T$_2$  &$\surd$ & $\surd$    \\
     513   &  T$_1$  &         &           & \hskip 0.7in      1159     &  T$_1$  &        &            \\
     528   &  T$_2$  &$\surd$  & $\surd$   & \hskip 0.7in      1163     &  A$_2$  &        &            \\
     558   &  T$_1$  &         &           & \hskip 0.7in      1166     &  T$_1$  &        &            \\
     573   &  E      &         & $\surd$   & \hskip 0.7in      1175     &  A$_2$  &        &            \\
     589   &  T$_2$  &$\surd$ & $\surd$    & \hskip 0.7in      1189     &  A$_1$  &        & $\surd$    \\
     592   &  T$_1$  &        &            & \hskip 0.7in      1198     &  T$_1$  &        &            \\
     605   &  T$_2$  &$\surd$ & $\surd$    & \hskip 0.7in      1206     &  E      &        &$\surd$     \\
     612   &  E      &        & $\surd$    & \hskip 0.7in      1207     &  T$_2$  &$\surd$ & $\surd$    \\
     636   &  T$_2$  &$\surd$ &$\surd$     & \hskip 0.7in      1232     &  T$_2$  &$\surd$ & $\surd$    \\
     637   &  T$_1$  &        &            & \hskip 0.7in      1237     &  T$_1$  &        &            \\
     640      &  E      &        &$\surd$     & \hskip 0.7in      1271     &  T$_1$  &        &            \\
     653      &  T$_2$  &$\surd$ &$\surd$     & \hskip 0.7in      1288      &  T$_2$  &$\surd$ & $\surd$    \\
     664      &  E      &        &$\surd$     & \hskip 0.7in      1299      &  T$_1$  &        &            \\
     675      &  A$_1$  &        &$\surd$     & \hskip 0.7in      1304      &  E      &        & $\surd$    \\
     677      &  T$_1$  &        &            & \hskip 0.7in      1356      &  T$_1$  &        &            \\
     710      &  T$_2$  &$\surd$ &$\surd$     & \hskip 0.7in      1356      &  T$_2$  &$\surd$ & $\surd$    \\
     714      &  E      &        &$\surd$     & \hskip 0.7in      1365      &  E      &        & $\surd$    \\
     715      &  A$_1$  &        & $\surd$    & \hskip 0.7in      1395      &  A$_1$  &        & $\surd$    \\
     723      &  T$_1$  &        &            & \hskip 0.7in      1420      &  T$_2$  &$\surd$ & $\surd$    \\
     726      &  T$_2$  &$\surd$ &$\surd$     & \hskip 0.7in      1432      &  T$_2$  &$\surd$ & $\surd$    \\
     740      &  T$_2$  &$\surd$ &$\surd$     & \hskip 0.7in      1455      &  T$_2$  &$\surd$ & $\surd$    \\
     787      &  T$_1$  &        &            & \hskip 0.7in      1456      &  E      &        & $\surd$    \\
     792      &  E      &        &$\surd$     & \hskip 0.7in      1464      &  A$_1$  &        & $\surd$    \\
     795      &  A$_1$  &        &$\surd$     & \hskip 0.7in                    &         &        &     \\
\end{tabular}
\end{ruledtabular}
\end{table*}

 The density of states (DOS) and optical spectra of the \BNS cluster 
calculated within DFT are shown in Fig. \ref{optical}.
The HOMO-LUMO gap of this cluster is 5.0 eV. The HOMO has a predominant N character whereas the
LUMO is delocalized over the whole cluster. The states near the HOMO and the LUMO have more N character
than B character. The HOMO has $t_2$ symmetry and the LUMO has $a_1$ symmetry.
The HOMO-LUMO transition is weak though symmetry allowed.
The optical spectra shows a sharp peak at 5.4 eV which occurs from transitions from occupied states
within 0.59 eV of HOMO to the unoccupied states within 0.43 eV of LUMO. This peak has contributions
from the $t_2 \rightarrow t_2$ and $e \rightarrow t_2$  transitions. It also has weak contributions from
$t_2 \rightarrow e$ transitions.
There are other higher energy peaks  seen at 9.7 eV and 13.8 eV. The small peak at 9.7 eV arises
mainly from a $t_2 \rightarrow t_2$ transition involving occupied states lying 3.4 eV below
HOMO and unoccupied
states 1.3 eV above the LUMO. The broad peak at 13.8 eV is due to a large number of weak transitions
involving a large number of low lying states rather than any sharp strong transition.
It should be noted that the present calculations are performed within the PBE-GGA at zero temperature
and hence the predicted optical spectra will be useful only for qualitative comparison. More accurate
spectra can be obtained by the TDDFT or GW approximation.

\begin{figure}
\epsfig{file=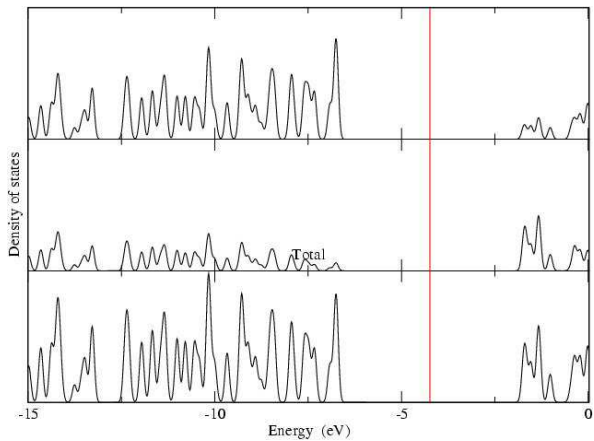,width=0.9\linewidth,clip=true}
\caption{ The density of states  of \BNS  cluster.}
\label{spn}
\end{figure}

\begin{figure}
\epsfig{file=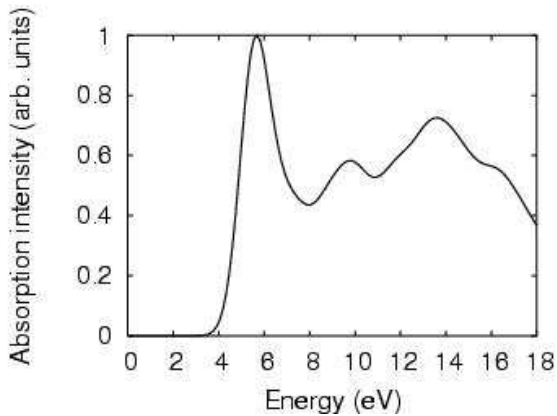,width=0.9\linewidth,clip=true}
\caption{ The optical absorption spectra of \BNS  cluster.}
\label{optical}
\end{figure}

   In summary, density functional theory at the level of PBE-GGA is used to examine 
the vibrational stability of the proposed fullerene-like cage structure of the \BNS
observed in a recent electron-radiation measurement. The cage structure was proposed on the criteria
that it satisfied the isolated six square rule for alternate BN cages and has dimension
similar  to  the single-shell  BN cages observed in the HRTEM images. The calculations 
indicate somewhat smaller dimension ($\sim 8 $ \AA) of the \BNS cage than that those 
observed in experiments ($9-10$ \AA).  The cage is energetically and vibrationally stable.
It has vertical and adiatbatic  ionization potential of 8.2 and  8 eV, respectively. 
Its mean static dipole polarizability is  79 \AA$^3$.  We have predicted IR,  Raman and 
optical spectra at the DFT GGA level. 
We hope that the predicted spectra in this work will stimulate experimental 
measurements of these spectra. The measured spectra, in combination with 
the predicated spectra in this work, will play a decisive role in confirmation of the
proposed candidate structure.

        The Office of Naval Research, directly and through the Naval Research Laboratory, and and the 
Department of Defense's   High Performance Computing Modernization Program, through the Common High 
Performance Computing Software Support Initiative Project MBD-5, supported this work.

\end{document}